\documentclass[runningheads]{llncs}
\usepackage[T1]{fontenc}
\usepackage{amsmath}
\usepackage{hyperref}
\hypersetup{colorlinks=true}
\usepackage{graphicx,verbatim}

\usepackage{colortbl}
\usepackage{color}
\usepackage{multirow}
\usepackage{booktabs}
\usepackage{enumitem}
\usepackage{caption}   %
\usepackage{amssymb}
\usepackage{subcaption}  %
\usepackage{graphicx}
\usepackage[table,xcdraw]{xcolor}
\begin{document}

\title{Downstream Analysis of Foundational Medical Vision Models for Disease Progression}

\titlerunning{Foundational Models for Disease Progression}

\author{Ba\c{s}ar Demir$^1$ \and Soumitri Chattopadhyay$^1$ \and Thomas Hastings Greer$^1$ \and \\ Boqi Chen$^1$ \and Marc Niethammer$^2$}  %
\authorrunning{Demir et al.}
\institute{University of North Carolina at Chapel Hill \and University of California, San Diego \\
    \email{\{bdemir, soumitri, tgreer, bqchen\}@cs.unc.edu} \\
    \email{mniethammer@ucsd.edu}}

\maketitle              %

\begin{abstract}
Medical vision foundational models are used for a wide variety of tasks, including medical image segmentation and registration. This work evaluates the ability of these models to predict disease progression using a simple linear probe. We hypothesize that intermediate layer features of segmentation models capture structural information, while those of registration models encode knowledge of change over time. Beyond demonstrating that these features are useful for disease progression prediction, we also show that registration model features do not require spatially aligned input images. However, for segmentation models, spatial alignment is essential for optimal performance. Our findings highlight the importance of spatial alignment and the utility of foundation model features for image registration.


\end{abstract}

\section{Introduction}
Deep learning methods have emerged as powerful tools in the medical imaging computing for disease diagnosis~\cite{khan2021machine,sharma2022deep}, segmentation~\cite{wang2022medical,hesamian2019deep}, and registration~\cite{yang2017quicksilver,balakrishnan2019voxelmorph}. A critical application is the early prediction of disease progression, enabling timely interventions and personalized treatment planning. %

Longitudinal imaging data provides an opportunity to capture disease progression over time. %
Several public longitudinal datasets are available. For example, from the Osteoarthritis Initiative (OAI)~\cite{nevitt2006osteoarthritis}  or from  the Alzheimer's Disease Neuroimaging Initiative (ADNI)~\cite{jack2008alzheimer}. However, disease progression models generally need to be trained for every application.%

In recent years, foundation models have gained traction for  segmentation ~\cite{sammed3d,swinunetr} and registration~\cite{tian2024unigradicon,demir2024multigradicon}. These models, trained on diverse and large-scale datasets, have demonstrated strong generalization capabilities. Foundation model features have been utilized for diverse downstream tasks~\cite{sun2024foundation,wang2023real}; however, to the best of our knowledge, they have not been explored for disease progression analysis. In this study, we investigate whether pretrained foundation model features contain informative representations for disease progression using the OAI knee dataset~\cite{nevitt2006osteoarthritis}. As is common in representation learning, we use a linear probing approach to assess the suitability of a representation. 

\textbf{Our contributions are as follows:}

\begin{itemize}
    \item[1)] We present the first downstream analysis of foundational medical vision models for disease progression prediction; 
    \item[2)] We conduct extensive experiments on the effects of preprocessing/layer selection strategies on downstream performance; 
    \item[3)] We compare the performance of foundation models to specialized models.
\end{itemize}

\section{Related Work}
Assessing progression requires longitudinal data. Such datasets typically consist of multiple imaging modalities, clinical measures, and demographics~\cite{jack2008alzheimer,nevitt2006osteoarthritis}. 

In the field of knee osteoarthritis (OA) diagnosis and progression prediction, various methods have been proposed to predict the current disease state and its future development. While some approaches rely solely on clinical data~\cite{ningrum2021deep} or radiographs~\cite{abd2023fully}, others use a combination of both~\cite{guan2022deep}. Additionally, MRIs have been shown to contain valuable signals related to OA progression. Schiratti et al.~\cite{schiratti2021deep} utilized MRIs and clinical information to predict cartilage degradation over a 12-month period. Hu et al.~\cite{hu2023deepkoa} extended this approach by using slices from 3D MRIs, combined with clinical variables, to detect OA progression. Panfilov et al.~\cite{panfilov2022predicting} initially employed CNNs to process MRI slices and aggregated the extracted features using transformers for progression prediction. Their work was later extended to incorporate multiple MRI sequences, improving predictive performance. However, all these imaging-based methods require training for \emph{feature extraction}. \emph{In contrast, our work explores the possibility of performing these tasks without the need for feature extractor retraining.}

Medical imaging encompasses a wide range of modalities and predictive tasks. This diversity has driven the development of foundation models capable of handling multiple tasks across different domains without requiring retraining: e.g., segmentation models ~\cite{liu2022isegformer,liu2022pseudoclick,sammed3d,swinunetr} and registration models~\cite{tian2024unigradicon,demir2024multigradicon}. Since these models are trained on a diverse set of datasets, they are expected to provide strong feature extractors for downstream tasks. Despite their success in numerous applications, their potential in disease progression analysis remains unexplored. \emph{In this study, we investigate the effectiveness of foundation models in capturing disease progression.}

\section{Methodology}
Let $p_{i}[t]$ denote the 3D scan of the $i$-th patient at timepoint $t$. Our aim is to extract information about disease progression between two scans taken at different timepoints, specifically $p_{i}[t_{1}]$ and $p_{i}[t_{2}]$. To achieve this, we benefit from foundational registration and segmentation models. We hypothesize that the intermediate (bottleneck) outputs of these networks are effective representations to capture disease progression between two given scans.

\subsection{Affine Registration via Inverse Consistency by Construction}
It is often desirable to affinely register images to a template before nonparametric registration, because the affine component of the transform might be a nuisance variable, e.g., due to inconsistent patient positioning. In addition, for large-scale analysis (in our experiments over 20,000 patient-template registrations) fast affine registration is desirable. Prior work~\cite{shen2019networks} reported that conventional approaches perform poorly for cross-subject affine registration on the OAI data. We therefore develop a deep-learning approach building upon inverse consistent by construction affine layers~\cite{greer2023inverse} trained on the uniGradICON~\cite{tian2024unigradicon} model. The architecture proposed in \cite{greer2023inverse} performs affine and nonparametric registration via the architecture
\begin{equation}
    \texttt{TSC}\{\Psi_1, \texttt{TSC}\{\Psi_2, \texttt{TSC}\{\Xi_1,\Xi_2\}\}\}\,,
\end{equation}
where $\Psi_i$ are inverse consistent affine networks,  $\Xi_i$ are inverse consistent by construction nonparametric networks, and $\texttt{TSC}$ denotes the inverse consistent composition operator defined in~\cite{greer2023inverse} . To create an architecture that purely performs affine registration, we simply only use inverse consistent affine layers: 
\begin{equation}
    \texttt{TSC}\{\Psi_1, \texttt{TSC}\{\Psi_2, \texttt{TSC}\{\Psi_3,\texttt{TSC}\{\Psi_4, \Psi_5\}\}\}\}\,.
\end{equation}

We train this affine model for 180 epochs at a low resolution of [88 x 88 x 88], and upsample the transforms to the original image resolution. The resulting registrations are accurate, achieving a Dice score of 40\% on average on femoral and tibial cartilage compared to 38\% for NiftyReg~\cite{shen2019networks}. %

\subsection{Registration Foundation Model Feature Extraction}
We use the uniGradICON~\cite{tian2024unigradicon} registration model as our feature extractor. Its architecture consists of TwoStep ($\texttt{TS})$ and DownSample ($\texttt{DS}$) operators~\cite{tian2023gradicon}:
\begin{align}
    \texttt{TS}\{\Psi_\theta^1, \Psi_\theta^2\}[I^A, I^B] &:= \Psi_\theta^1[I^A, I^B] \circ \Psi_\theta^2[I^A \circ \Psi_\theta^1[I^A, I^B], I^B]\,,\\
    \texttt{DS}\{\Psi_\theta\}[I^A, I^B] &:= \Psi_\theta[\texttt{averagePool}(I^A, 2), \texttt{averagePool}(I^B, 2)]\,.
\end{align}
UniGradICON uses 4 U-Nets~\cite{cicek2016_3dunet} ($\Psi_\theta^i$) that predict displacement fields and constructs the registration network by stacking ($\texttt{TS})$ and ($\texttt{DS}$) operators as
$\Phi_\theta := \texttt{TS}\{\texttt{TS}\{\texttt{DS}\{\texttt{TS}\{\texttt{DS}\{\Psi_\theta^1\}, \Psi_\theta^2\}\}, \Psi_\theta^3\}, \Psi_\theta^4\}\,.$ The networks operate at $\frac{1}{4}$, $\frac{1}{2}$, $1$ and $1$ input resolution. In our experiments, we use the bottleneck feature maps of the U-Nets, i.e., the deepest latent representations within $\Psi_\theta^i$, as our feature embeddings. We hypothesize that these features capture high-level spatial correspondences and are therefore useful for downstream tasks.

\subsection{Segmentation Foundation Model Feature Extraction}
We use both foundation and task-specific segmentation models for feature extraction. On the foundational side, we utilize the pretrained image encoder of SAM-Med3D \cite{sammed3d} and the bottleneck features of the pretrained SwinUNETR \cite{swinunetr}. Additionally, we train a UNet \cite{cicek2016_3dunet} for cartilage and bone segmentation on the OAI-ZIB dataset \cite{ambellan2019automated}. We also extract the bottleneck features from this model.

\subsection{Training Details}
For all experiments, we use a simple one-layer linear classifier (i.e., a linear probe \cite{alain2016understanding}). Implemented in PyTorch \cite{pytorch}, we train this linear classifier for 10K iterations using AdamW~\cite{adamw}, with a learning rate of $10^{-3}$.

We also train a UNet for 50 epochs, using AdamW~\cite{adamw} with a learning rate of $10^{-3}$, for femur, tibia, femoral cartilage and tibial cartilage segmentation. This UNet has channel dimensions of (8, 16, 16, 32, 32, 64, 64, 128, 128), stride values of (2, 1, 2, 1, 2, 1, 2, 1), and a total of 9 residual units. The implementation is based on the MONAI framework \cite{cardoso2022monai}. 
        
\section{Results}
We assess the performances of the foundational registration network uniGradICON, the foundational segmentation models SAM-Med3D~\cite{sammed3d},~SwinUNETR \cite{swinunetr}; and of  a UNet which was specifically trained for knee bone and cartilage segmentation on the OAI dataset \cite{nevitt2006osteoarthritis}. We keep the experimental setup simple and fair across methods by always using a simple one-layer linear classifier that directly maps features to the predicted classes.

\noindent \textbf{Dataset.} We evaluate our approach using a subset of the OAI dataset \cite{nevitt2006osteoarthritis}, which includes 2,244 patients and their longitudinal MRI scans of the left and right knees over a period of up to 96 months. We use five timepoints for our prediction tasks, as MRIs are unavailable at the 60-month timepoint. Patients are randomly split into training (50\%), validation (12.5\%), and testing (the remainder) sets.  
The dataset contains WOMAC pain scores and KL grades, where the labels range from 0 to 20 and 0 to 4, respectively, as well as medial joint space width (JSW), which represents the distance between the femur and tibia. For pain prediction, we define WOMAC~$<5$ as ``no pain'' and $\geq 5$ as ``pain.'' For KLG, we merge KLG~$=0$ and KLG~$=1$, as osteoarthritis is considered definitive only when KLG~$\geq 2$ \cite{kohn2016classifications}.\\

\noindent \textbf{General Questions.} We hypothesize that registration features are informative about disease progression, as they are trained to find correspondences between images without requiring labels (e.g., segmentations). We seek to answer the following questions: 1) How does the performance in predicting pain scores for the registration model and a given reference healthy atlas compare to the segmentation model encoders? (Sec. \ref{sec:prediction}); 2) Do registration/segmentation features capture disease progression for images from two time points? (Sec. \ref{sec:progress}); 3) Can registration/segmentation features be used to predict the future disease condition 24 months ahead? (Sec. \ref{sec:predict24}).\\

\noindent \textbf{Experimental Design.} We investigate several factors that might affect the standardization of extracted features and, consequently, the performance of downstream classifications. To this end, we apply different combinations of image preprocessing steps, as follows:

\noindent \textbullet \textbf{ Affine Atlas Alignment:} We affinely align input images to a atlas to remove the affine component of deformation.%

\noindent \textbullet \textbf{ Nonparametric Registration to the Atlas:} We apply this step only to segmentation networks to examine whether they learn from texture or truly capture anatomical differences.

\noindent \textbullet \textbf{ ROI Cropping:} We define a region of interest (ROI) focusing on the joint and surrounding bone structures while excluding muscles and unrelated anatomical components. This ensures  features extraction from the primary area of interest.

These preprocessing steps aim to standardize input representations and enhance the effectiveness of foundation model features in predicting OA-related clinical scores or making comparisons across images.\\

\noindent \textbf{Atlas Generation.} We generate an atlas image using enrollment scans of 200 healthy patients with a KL grade of 0 and a WOMAC score of 0. This atlas serves as a reference for the registration network, enabling it to assess how far a given knee deviates from a healthy one when needed. To generate the atlas, we use the template normalization algorithm of \cite{avants2008symmetric}. This algorithm can be applied with any registration backbone: we use uniGradICON as the registration algorithm.%

\subsection{KL Grade Prediction}
\label{sec:prediction}

\begin{table}[t]
\caption{KL Grade Classification Accuracy and Area Under the Curve (AUC). "A" denotes affine alignment, "D" denotes nonparametric alignment, and "C" denotes ROI cropping.}
\label{tab:klg-prediction}
\setlength{\tabcolsep}{5pt}
\renewcommand{\arraystretch}{1.2}
\resizebox{\columnwidth}{!}{%
\begin{tabular}{ccc|ccccccc}
\hline
\multicolumn{3}{c|}{\textbf{Preproc.}} & \textbf{FM-Reg.}     & \multicolumn{4}{c}{\textbf{FM-Seg.}}                                            & \multicolumn{2}{c}{\textbf{Specialist-Seg.}} \\ \cline{4-10} 
\textbf{A}  & \textbf{D}  & \textbf{C} & \textbf{uniGradICON} & \multicolumn{2}{c}{\textbf{SAM-Med3D}} & \multicolumn{2}{c}{\textbf{SwinUNETR}} & \multicolumn{2}{c}{\textbf{UNet}}            \\ \hline
            &             &            & \textit{w/ atlas}    & \textit{w/o atlas} & \textit{w/ atlas} & \textit{w/o atlas} & \textit{w/ atlas} & \textit{w/o atlas}    & \textit{w/ atlas}    \\ \hline
\rowcolor[HTML]{DAE8FC} 
            &             &            & 63.4, 84.8           & 57.0, 79.4         & 56.8, 79.0        & 54.9, 74.8         & 53.6, 74.7        & 45.5, 61.7            & 41.3, 53.7           \\
\checkmark  &             &            & 62.7, 84.0           & 62.3, 83.7         & 62.3, 83.8        & 59.0, 80.0         & 58.2, 79.8        & 48.4, 64.4            & 43.2, 63.8           \\
\rowcolor[HTML]{DAE8FC} 
\checkmark  &             & \checkmark & 65.9, 86.8           & 65.7, 86.7         & 66.3, 86.6        & 62.0, 84.1         & 60.9, 83.1        & 37.9, 62.5            & 44.8, 59.7           \\
            & \checkmark  &            & -                    & 62.3, 83.0         & 62.2, 82.8        & 60.8, 81.4         & 62.1, 81.6        & 48.1, 69.4            & 46.5, 69.3           \\
\rowcolor[HTML]{DAE8FC} 
            & \checkmark  & \checkmark & -                    & 66.3, 86.4         & 65.9, 86.1        & 65.4, 85.8         & 62.5, 83.6        & 50.5, 69.9            & 48.6, 70.4           \\ \hline
\end{tabular}%
}
\end{table}

We explore the use of foundation model features to predicting KL grade. We provide a healthy atlas as a reference to uniGradICON, using it as a fixed input for all feature extraction operations. This allows the model to compare the given moving image with the healthy reference and determine the degree of deviation from the healthy knee. For segmentation networks, we investigate two input strategies: using image features alone and computing the difference between image and atlas features. Additionally, we explore various preprocessing steps to better understand the key factors influencing feature extraction quality.

\textbf{Affine alignment is essential for segmentation models.} For non-aligned images, Table~\ref{tab:klg-prediction} shows that uniGradICON features outperform the best performing segmentation model (SAM-Med3D without atlas) by 6.4\% in accuracy and 5.4\% in area under the curve (AUC) scores. After affine alignment, we observe a 2–5\% increase in accuracy and a 4–10\% increase in AUC. With this improvement, SAM-Med3D narrows its performance gap to 0.4\% in accuracy and 0.3\% in AUC. However, registration features maintain comparable performance before and after alignment, demonstrating their robustness to non-aligned inputs.

\textbf{Segmentation models learn from the texture.} To determine whether segmentation models capture structural differences, such as thinner cartilage or reduced joint space width, we nonparametrically register images to a healthy atlas. This process helps mitigate anatomical variations across images, making each image structurally resemble the atlas while preserving textural differences. Table \ref{tab:klg-prediction} shows that the results obtained using only affine registration and only nonparametric registration are highly similar, with performance differences generally around or below 1\%. This suggests that segmentation model features primarily capture textural differences rather than anatomical variations.

\textbf{Good segmentation performance does not imply good features.} We additionally trained a deep U-Net for bone and cartilage segmentation which reaches $\sim$0.98\% bone and $\sim$0.84\% cartilage segmentation Dice scores. On the other hand, the other segmentation networks were never trained on this dataset. However, Table \ref{tab:klg-prediction} shows that this U-Net results in the worst performance with $\sim$10\% lower AUC and accuracy scores on all scenarios. This indicates that the feature usefulness does not directly correlate to segmentation performance. Note that it is possible to make different architectural designs/intermediate feature selections, but for this study, we always use the bottleneck layers for simplicity. %

\begin{figure}[t]
    \centering
    \begin{subfigure}[b]{0.49\textwidth}
        \includegraphics[width=\textwidth]{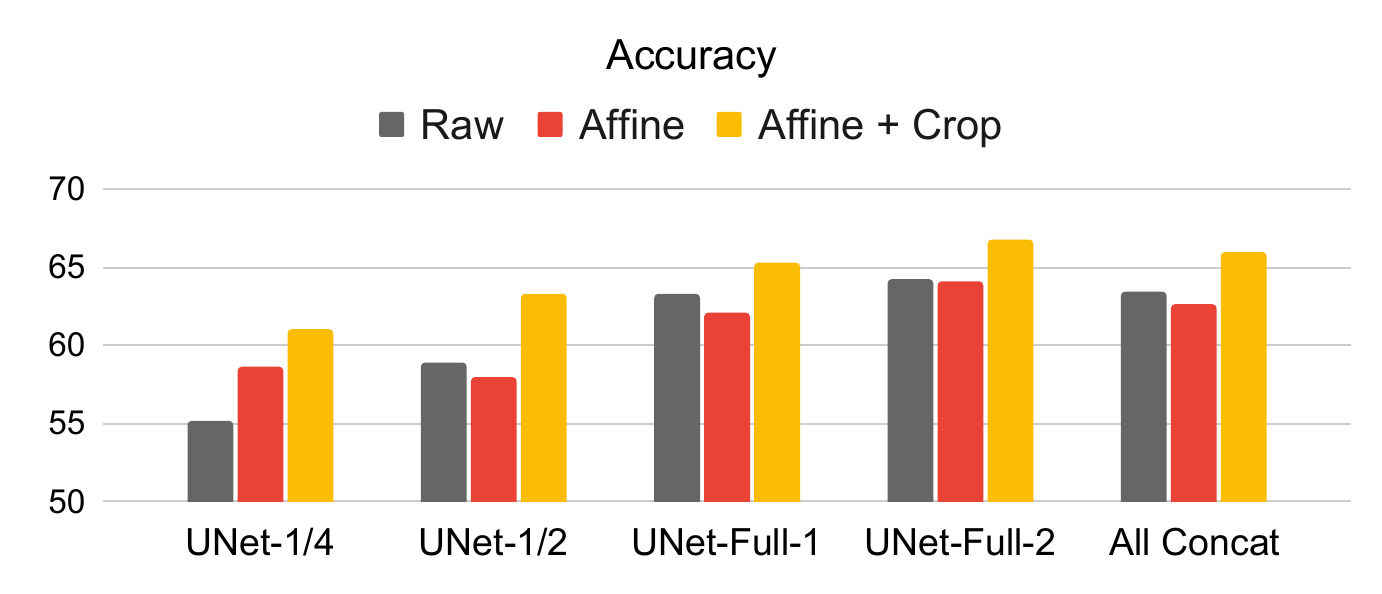}
        \label{fig:image1}
    \end{subfigure}
    \hfill
    \begin{subfigure}[b]{0.49\textwidth}
        \includegraphics[width=\textwidth]{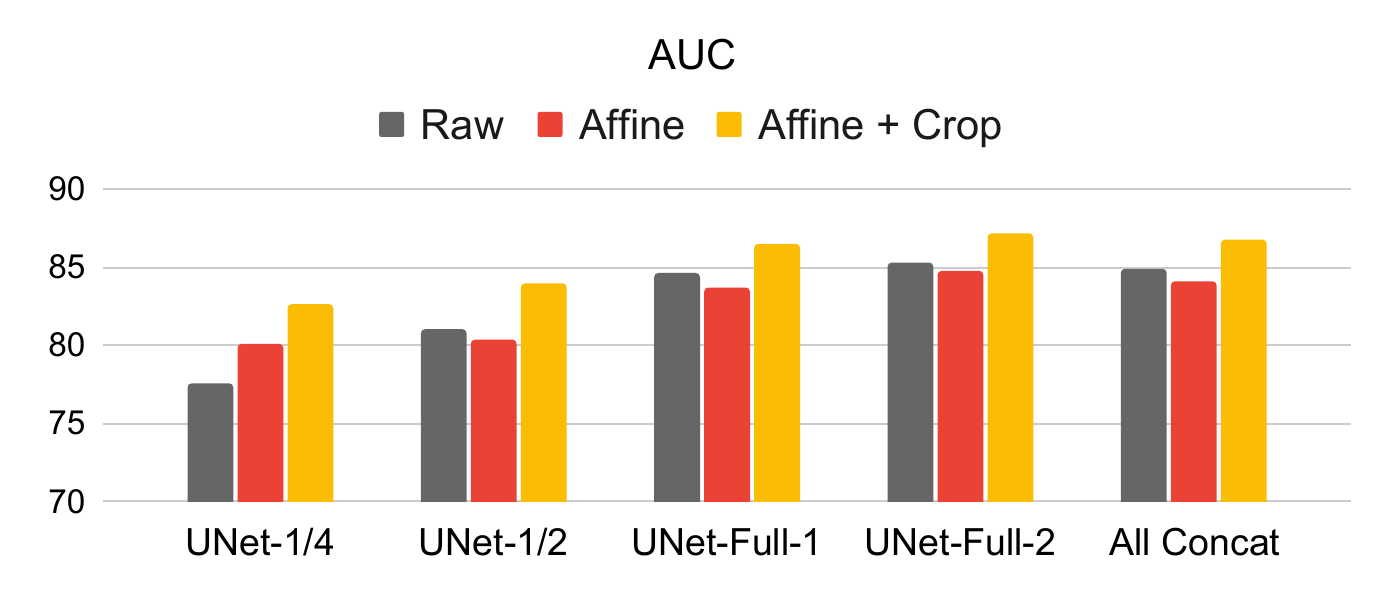}
        \label{fig:image2}
    \end{subfigure}
    \caption{Performance of bottleneck features from sub-networks of uniGradICON using different preprocessing techniques. We observe that later layers of the networks are more robust to misaligned inputs, and cropping  improves performance.}
    \label{fig:unigradicon}
\end{figure}

\textbf{ROI cropping increases the performance.} Knee OA affects the knee joint, however MRIs of the knee cover a larger area. Hence, we investigate the effect of cropping, so that extracted features focus on the region of interest rather than including features from other parts of the knee. Table~\ref{tab:klg-prediction} demonstrates that cropping always improves the performance with respect to both accuracy and AUC. For instance, uniGradICON with cropping improves accuracy by $\sim$3\% and AUC by $\sim$2.8\%;  for SwinUNETR w/o atlas and nonparametric preprocessing, cropping improves accuracy by $\sim$4.6\% and AUC by $\sim$4.4\%.

\textbf{Taking the difference from the atlas features does not have a significant effect.} We aim to further guide the classification based on segmentation features by incorporating extracted atlas features. We hypothesize that if the difference between the query image features and the healthy atlas features is small, this might indicate that the query is healthy. However, we do not observe a significant effect. We obtain on-par performances in both cases.

\textbf{Later uniGradICON networks are more informative for progression.} We perform an ablation study on the performance of different uniGradICON subnetworks. Figure~\ref{fig:unigradicon} shows that later full resolution layers are more robust to alignment compared to initial layers. This makes sense since initial layers operate at lower resolution and may mainly be responsible for affine alignment, so that later layers are not affected by non-aligned input images. Furthermore, we observe that ROI cropping increases the classification scores for every layer.

\begin{table}[t]
\centering
\caption{Classification results for disease progression from two timepoints.}
\setlength{\tabcolsep}{5pt}
\label{tab:compare}
 \resizebox{\columnwidth}{!}{
\begin{tabular}{ccccccccc} 
\hline
                                   & \multicolumn{2}{c}{\textbf{uniGradICON}} & \multicolumn{2}{c}{\textbf{SAM-Med3D}} & \multicolumn{2}{c}{\textbf{SwinUNETR}} & \multicolumn{2}{c}{\textbf{UNet}} \\ \cline{2-9} 
\multirow{-2}{*}{\textbf{Config.}} & KLG                 & JSW                & KLG                & JSW               & KLG                & JSW               & KLG             & JSW             \\ \hline
\textbf{w/o Affine}                &                     &                    &                    &                   &                    &                   &                 &                 \\
\rowcolor[HTML]{DAE8FC} 
ACC                                & 88.0                & 72.0               & 88.0               & 70.2              & 88.6               & 70.4              & 85.1            & 40.3            \\
AUC                                & 71.9                & 67.4               & 64.7               & 63.1              & 59.5               & 61.1              & 52.5            & 55.3            \\
\rowcolor[HTML]{DAE8FC} 
F1                                 & 61.4                & 61.0               & 56.3               & 58.8              & 51.6               & 56.4              & 49.7            & 40.3            \\ \hline
\textbf{w/Affine}                  &                     &                    &                    &                   &                    &                   &                 &                 \\
\rowcolor[HTML]{DAE8FC} 
ACC                                & 87.8                & 71.4               & 88.3               & 73.7              & 87.2               & 73.2              & 86.8            & 43.2            \\
AUC                                & 71.8                & 66.5               & 71.6               & 67.9              & 62.2               & 64.1              & 59.4            & 53.5            \\
\rowcolor[HTML]{DAE8FC} 
F1                                 & 60.8                & 60.7               & 59.0               & 61.9              & 57.2               & 59.7              & 49.0            & 42.7            \\ \hline
\end{tabular}%
}
\end{table}

\subsection{Progression Prediction For A Given Scan Pair}
\label{sec:progress}
In this setup, rather than directly predicting scores from a given image, we evaluate model performance in detecting disease progression. Given an enrollment image and a follow-up image from any two time points, we aim to determine whether OA progression has occurred. For the registration network, we directly input both images and extract bottleneck features. In contrast, for segmentation-based features, we process both images through the network, extract their respective features, and concatenate them.

Since OA progression can be defined in multiple ways~\cite{martel2023magnetic}, we adopt two of the most widely used criteria: a change in KL grade~\cite{zhang2014development} or a reduction of at least 0.5mm in the minimum medial joint space width (JSW) over a period of at least 12 months~\cite{bruyere2005three}.

To evaluate our models, we use two sets of experiments: one with affine pre-alignment and one without. Table~\ref{tab:compare} shows that uniGradICON outperforms other comparison networks in both AUC and F1 score across both experimental setups (KLG/JSW) without pre-affine alignment. Specifically, it performs ~7\% better in AUC and ~5\% better in F1 score than the closest performing method, SAM-Med3D. However, we observe a consistent performance improvement in segmentation networks when affine alignment is applied, with SAM-Med3D achieving nearly on-par results with uniGradICON and slightly outperforming uniGradICON on the JSW setup. In contrast, the SwinUNETR and UNet models exhibit significantly lower performance compared to SAM-Med3D and uniGradICON. Hence, we can state that registration features are more robust to alignment changes for the disease progression task, and a good segmentation model can perform comparably after an additional alignment stage.

\subsection{Progression Prediction 24 Months in Advance}
\label{sec:predict24}
Here, we use foundation model features for disease progression prediction, aiming to predict KL grade and WOMAC score 24 months in beforehand. We compare predictions for month 96 using different time intervals for prior images. All images are affinely aligned to the atlas before extracting and concatenating segmentation features. For the registration network, we use the atlas as input and concatenate features from given months. As a baseline, we compare with uniMMMV-Th.~\cite{chen2024unified}, trained on 2D cartilage thickness maps from 3D MRIs.

Table \ref{tab:MMMV} demonstrates that uniMMMV-Th. outperforms other classifiers on AP. We believe this gap arises primarily because uniMMMV-Th. is trained on thickness maps, which are powerful biomarkers for OA detection. Among all feature extractors, SAM-Med3D achieved the highest performance, surpassing others by 2.3\% in KL grade accuracy, 3.2\% in AUC, and 5.4\% in AP for predictions using a single image from month 72. It also outperforms uniMMMV-Th. by 4.1\% in KL grade accuracy, 27\% in WOMAC classification, and 0.4\% in WOMAC AUC. Additionally, while uniMMMV-Th. improves with more time points, our linear probing approach struggles due to architectural limitations.

\begin{table}[t]
\centering
\caption{96th-Month KLG and WOMAC Prediction Results. "ACC" denotes accuracy, "AUC" denotes area under the curve, and "AP" denotes average precision.}
\label{tab:MMMV}
\setlength{\tabcolsep}{3.5pt}
\renewcommand{\arraystretch}{1}
\resizebox{\textwidth}{!}{%
\begin{tabular}{ccccccccccc}
\hline
\textbf{Months}  & \multicolumn{2}{c}{\textbf{uniGradICON}} & \multicolumn{2}{c}{\textbf{SAM-Med3D}} & \multicolumn{2}{c}{\textbf{SwinUNETR}} & \multicolumn{2}{c}{\textbf{UNet}} & \multicolumn{2}{c}{\textbf{uniMMMV-Th.}} \\ \cline{2-11} 
\textbf{of data} & KLG                & WOMAC               & KLG               & WOMAC              & KLG               & WOMAC              & KLG             & WOMAC           & KLG                & WOMAC               \\ \hline
\multicolumn{2}{l}{\textbf{Only 72}}  &                     &                   &                    &                   &                    &                 &                 &                    &                     \\
\rowcolor[HTML]{DAE8FC} 
ACC              & 78.5               & 81.2                & 81.1              & 82.7               & 78.8              & 81.3               & 72.3            & 58.2            & 77.0               & 55.7                \\
AUC              & 80.9               & 62.6                & 84.1              & 64.2               & 75.7              & 62.6               & 56.7            & 52.9            & 87.3               & 63.8                \\
\rowcolor[HTML]{DAE8FC} 
AP               & 39.6               & 8.8                 & 45.0              & 8.7                & 36.9              & 8.3                & 27.8            & 8.2             & 56.4               & 26.3                \\ \hline
\multicolumn{2}{l}{\textbf{24 -- 72}} &                     &                   &                    &                   &                    &                 &                 &                    &                     \\
\rowcolor[HTML]{DAE8FC} 
ACC              & 79.2               & 82.6                & 79.9              & 83.6               & 79.1              & 81.2               & 67.3            & 68.9            & 79.9               & 56.7                \\
AUC              & 72.4               & 50.3                & 83.4              & 60.0               & 75.2              & 54.2               & 54.9            & 51.1            & 88.0               & 64.0                \\
\rowcolor[HTML]{DAE8FC} 
AP               & 36.6               & 8.2                 & 43.5              & 8.3                & 36.5              & 8.1                & 26.9            & 8.0             & 56.3               & 27.1                \\ \hline
\multicolumn{2}{l}{\textbf{0 -- 72}}  &                     &                   &                    &                   &                    &                 &                 &                    &                     \\
\rowcolor[HTML]{DAE8FC} 
ACC              & 78.1               & 81.8                & 80.6              & 83.7               & 79.5              & 81.5               & 66.5            & 72.0            & 80.4               & 57.5                \\
AUC              & 69.5               & 52.9                & 82.5              & 57.6               & 68.3              & 51.9               & 66.5            & 51.0            & 88.0               & 63.8                \\
\rowcolor[HTML]{DAE8FC} 
AP               & 35.7               & 10.5                & 43.3              & 8.7                & 35.2              & 8.0                & 26.5            & 8.0             & 56.9               & 26.8                \\ \hline
\end{tabular}%
}
\end{table}

\section{Conclusion}
In this work, we conducted  downstream analyses for disease progression prediction using foundational registration, segmentation, and specialist segmentation network features. We observed that registration features can be effective for progression prediction even without affine pre-alignment of images, whereas affine alignment is important for segmentation networks; specialist segmentation networks do not provide good features for progression prediction. For future work, we aim to explore feature extraction layer selection strategies; more advanced architectures beyond a linear classifier; developing effective methods forfinetuning these models; and evaluating these approaches on additional datasets.

\section{Acknowledgements}
This research was, in part, funded by the National Institutes of Health (NIH) under other transactions 1OT2OD038045-01 and NIAMS 1R01AR082684. The views and conclusions contained in this document are those of the authors and should not be interpreted as representing official policies, either expressed or implied, of the NIH.

\bibliographystyle{splncs04}
\bibliography{mybibliography}

\end{document}